\documentstyle[twoside,fleqn,espcrc2]{article}
\title{Renormalization group and continuum limit in Quantum Mechanics}
\author{Janos Polonyi\address{Department of Physics,
   University Louis Pasteur,\\ Strasbourg, France and
   Department of Atomic Physics, L. E\"otv\"os University, \\ Budapest,
   Hungary}}
\begin{document}
\begin{abstract}
The running coupling constants are introduced in Quantum
Mechanics and their evolution is described by the help of the
renormalization group equation.
\end{abstract}
\maketitle
\section{INTRODUCTION}
The typical trajectories of the path integral for a free nonrelativistic
particle are nowhere differentiable \cite{fehib}. In fact, the dominant
contributions to the amplitude
\begin{equation}
\prod_k\int dx_ke^{{im\over2\hbar\Delta t}\sum_k\Delta x_k^2},
\end{equation}
$\Delta x_k=x_{k+1}-x_k$,
is such that the contributions to the sum in the exponent is order one,
i.e. ${\Delta x\over\Delta t}=O(\sqrt{1\over\Delta t})$. In other words,
the determination of the velocity within the time span $\Delta t$ yields
fluctuations $O(\sqrt{1\over\Delta t})$. The intrinsic `disorder' which
might be interpreted as an evidence of the fractal structure of the
trajectories represents the core of Quantum Physics for the canonical
commutation relations would be lost with
${\Delta x\over\Delta t}=o(\sqrt{1\over\Delta t})$. Furthermore, this
non-differentiability of the trajectories is the source of the
It\^o calculus \cite{ito} in Quantum Mechanics. One expects that
the velocity dependent interactions become more important for short
time processes. Our goal is to see whether the concept of running coupling
constant is meaningful in Quantum Mechanics and it shows
an enhancement for velocity dependent interactions at high energy
or short time.

Quantum Mechanics can formally be considered as a lattice
regulated Quantum Field Theory
in 0+1 dimension with lattice spacing $\Delta t$.
One finds that the transition amplitudes are not necessarily
ultraviolet finite when velocity dependent interactions are present. The power
counting argument shows that the vertex $({\Delta x\over\Delta t})^dx^s$ is
renormalizable, i.e. its contributions do not become more and more
ultraviolet singular in the higher orders of the perturbation expansion
for $d\le2+s$ \cite{pol}.
We shall find that some of the ultraviolet singularities predicted by the
simple power counting argument are curiously enough cancel in the simple
perturbation expansion and are only present when the expansion
is made around non-static trajectories. This raises the question about the
equivalence of the operator and the path integral formalism. These questions
will be considered here in the framework of the perturbation expansion.

The renormalization of Quantum Mechanics has
already been discussed in the presence of singular potential in \cite{rqm}.
Our motivation to look for the running coupling constant is different,
the goal is now to understand the time dependence of the transition amplitude
\begin{equation}
<x|e^{-{i\over\hbar}tH}|y>=<x|U(t)|y>=e^{{i\over\hbar}S(x,y;t)}
\end{equation}
for general regular interaction \cite{pol}.

\section{RUNNING COUPLING CONSTANTS}

The running coupling constants, $g_{ds}(t)$, are defined in the spirit of the
Landau-Ginsburg expansion by
\begin{equation}
S(x,y;t)=\sum_{ds}g_{ds}(t)(x-y)^d(x+y)^s,
\end{equation}
and they characterize the transition probabilities.
The vector indices are suppressed here for the sake of simplicity.
The coupling constants of the lagrangian is
$g_{ds}=t^{-2}\tilde g_{ds}$ after the separation of the time dimension
due to the cut-off $\Lambda=2\pi/\Delta t$ and $\hbar$.

The behavior of these effective coupling constants is
rather nontrivial even for a harmonic oscillator. The running mass
and frequency, given by
\begin{equation}
m(t)=2tg_{20}(t)={tm(0)\omega_0(1+\cos\omega(0)t)\over\sin\omega(0)t}
\end{equation}
and
\begin{equation}
\omega^2(t)={8g_{02}(t)\over tm(t)}
={4\over t^2}{1-\cos\omega(0)t\over1+\cos\omega(0)t}
\end{equation}
show periodicity in real time, a characteristic feature
for equidistant energy spectrum.

\section{RENORMALIZATION GROUP EQUATION}

The renormalization group equation is given by
\begin{equation}
e^{{i\over\hbar}S(x,y;t_1+t_2)}=\int dz
e^{{i\over\hbar}S(x,z;t_1)}e^{{i\over\hbar}S(z,y;t_2)}.
\end{equation}
The decimation consists of choosing $t_1=t_2=t$ and
the linearization of the resulting blocking transformation $t\to t'=2t$
yields the tree level scaling law
\begin{equation}
\tilde g_{ds}(t')=\biggl({t'\over t}\biggr)^{2-d}\tilde g_{ds}(t).
\end{equation}
Thus the relevant or marginal pieces of the lagrangian are the kinetic
energy, the vector and the scalar potential. Note that the
vertices with $2<d<2+s$ are renormalizable and irrelevant in the same time.
This results from the
loss of the relativistic invariance, i.e. the difference of the energy and the
inverse time dimensions. In fact, the former controls the ultraviolet
divergences while the latter is responsible for the scaling exponents
when the blocking is made in the time direction.

The one-loop evaluation of the differential renormalization group equation
in the limit $t_2\to0$ gives the Schr\"odinger equation for the logarithm
of the wave function. This method establishes a transparent relation
between the transition amplitude, $e^{-{i\over\hbar}S(x,y;t)}$,
and the hamiltonian, $H(t)=-{\hbar\over it}lnU(t)$.

\section{OPERATOR ORDERING AND RENORMALIZATION}

Eq. (7) gives the cut-off independence of the mass for $d=2$ and $s=0$. Another
less trivial implication of this equation is the requirement of
the midpoint prescription for the vector potential. The lagrangian
\begin{equation}
{m\over2}\biggl({\Delta x_k\over\Delta t}\biggr)^2+{\Delta x_k\over\Delta t}
A\biggl({x_{k+1}+x_k\over2}+\eta\Delta x_k\biggr),
\end{equation}
generates gauge dependent
amplitudes unless $\eta=0$ \cite{schul}. This quantum effect
results from the marginality of the term $\Delta x^2$ which  gives
the leading order $\eta$ dependence of the path integral.

Since the
saddle point of the integral in the renormalization group is different
for $t_1=t_2$ and $t_1\not=t_2\to0$ the one-loop scaling relations might well
be different for the lagrangian and the hamiltonian. It turns out that
the contribution to the renormalization group equations which is responsible
for the midpoint prescription is logarithmically
divergent in the blocking but remain finite in the framework of the
simple perturbation expansion or in the hamiltonian. In fact, the leading
order $\eta$ dependence of the path integral is proportional to
${\cal M}=\Delta t\sum_k<({\Delta x_k\over\Delta t})^2>_0$,
where $<\cdots>_0$ stands for the free expectation value.
Thus ${\cal M}=I\Lambda^{-1}$, where $I$ is a linearly
divergent loop integral, $I=A\Lambda+B\Lambda ln\Lambda$. In the expansion
around a static trajectory or in the derivation
of the hamiltonian the logarithmic correction
is absent. Thus the product ${\cal M}={1\over\Lambda}A\Lambda$ is finite,
in a manner which is reminiscent of the chiral anomaly.
The perturbation expansion
in the decimation is performed around a general, non-static trajectory
with time scale. This scale lets the usual logarithmic corrections to appear.
In another words,
a graph with zero primitive degree of divergence happened to
be finite in the expansion around the static trajectory. The hamilton formalism
based on such expansion is well defined without the introduction of the
ultraviolet cut-off. When the background trajectory is chosen to be
non-static the logarithmic divergence appears and makes regularization
necessary for the path integral.

The presence of the potentially divergent
but actually finite graph is reflected in the operator ordering problem
in Quantum Mechanics. The $\eta$ dependent vertices which are vanishing in
the classical limit i.e. for differentiable trajectories survive
the removal of the cut-off. This is in an apparent contradiction with the
result of Ref. \cite{riesz} where the renormalization of the
lattice regulated field theories was studied in the BPHZ scheme. The
finiteness of $\cal M$ in the framework of the simple perturbation expansion
makes renormalization unnecessary but leaves a non-uniformly convergent
loop integral in the theory. In the BPHZ program all graphs with nonpositive
primitive degree of divergence are subtracted and the convergence
of the manifestly convergent loop integrals is uniform. It remains to be
seen whether the operator ordering problem can change the renormalization
properties in Quantum Field Theory.

\section{EFFECTIVE LOW ENERGY THEORIES}

The renormalization group flow makes possible to construct low energy
effective models in Quantum Mechanics. According to the leading order
perturbation expansion the action
$S(x,y;t)={m(x-y)^2\over2t}-tU(x,y;t)$ where the potential $U(x,y;t)$
given by either
\begin{equation}
V\biggl({x+y\over2}\biggr),
\end{equation}
or
\begin{equation}
{1\over(d/2)!}\biggl({m\over3i\hbar}\biggr)^{d/2}
\biggl({x-y\over\sqrt{t}}\biggr)^dV\biggl({x+y\over2}\biggr),
\end{equation}
generates the same hamiltonian,
\begin{equation}
H={p^2\over2m}+V(x),
\end{equation}
for small but finite $t$. The potential (10) does
not follow the renormalized trajectory in the limit $t\to0$ and the
perturbation expansion breaks down for small enough $t$.
Thus the two potential contain the same relevant terms in the
hamiltonian continuum limit, (11).

\section{PATH INTEGRAL VERSUS OPERATOR FORMALISM}

The low energy effective models of Quantum Field Theory may contain
non-renormalizable operators. In this manner the class of effective theories
is larger than those of the fundamental, i.e. renormalizable theories.
For example the hadronic effective vertices which are described by multi-quark
operators are certainly generated in low energy QCD but can not be explicitely
present in the renormalizable QCD. Thus certain physically well
motivated vertices of the low energy theory can only be kept meaningful
in cut-off theories. As we insist on the removal of the cut-off then
these irrelevant operators must be eliminated from the bare lagrangian
and they can only be generated dynamically.

The effective model mentioned above shows the same features. The
scalar potential in (9) is well defined for arbitrary values of the cut-off.
The system (10) contains additional irrelevant terms which can not be
obtained directly from a hamiltonian which is given for $\Delta t=0$.
In this sense the cut-off version of Quantum Mechanics which is given
by the path integral in terms of the transition amplitudes for
infinitesimal time is more general than the operator formalism where
the equation of motion is a differential equation with the cut-off
already removed.

\end{document}